\documentclass[smallextended]{svjour3}
\usepackage{amssymb,amsmath,amscd}
\usepackage{amsfonts,bm,latexsym}

\smartqed
\journalname{GRG}

\renewcommand{\theequation}{\arabic{equation}}

\def\bes{\begin{subequations}}
\def\ees{\end{subequations}}
\def\be{\begin{equation}}
\def\ee{\end{equation}}
\def\bea{\begin{eqnarray}}
\def\eea{\end{eqnarray}}
\def\nn{\nonumber}
\def\p{\partial}
\def\mO{\mathcal{O}}

\def\tL{\widetilde{L}}
\def\rpm{\rho_{\pm}}
\begin{document}

\title{New forms and thermodynamics of the neutral rotating squashed black hole in
five-dimensional vacuum Einstein gravity theory}

\titlerunning{New forms and thermodynamics of the neutral rotating squashed black hole}

\author{Xiao-Dan Zhu$^1$ \and Di Wu$^1$ \and Shuang-Qing Wu$^{1,2}$ \and Shu-Zheng Yang$^1$}
\authorrunning{Zhu \and Wu \and Wu \and Yang}

\institute{\email{wdcwnu@163.com, Corresponding Author} \\
\email{sqwu@cwnu.edu.cn} \\
$^1$ College of Physics and Space Science, China West
Normal University, Nanchong, Sichuan 637002, People's Republic of China \\
$^2$ Institute of Theoretical Physics, China West Normal University,
Nanchong, Sichuan 637002, People's Republic of China}

\date{Received: date / Accepted: date}

\maketitle

\begin{abstract}
We initiated the program to look for new and simple forms for the five-dimensional rotating squashed
black holes by solving directly the equation of motion. In a recent paper, the metric ansatz of dimensional
reduction along the fifth spatial dimension was used to obtain a new but rather simple form for the
five-dimensional rotating uncharged black hole solution with squashed horizons via solving the vacuum
Einstein field equations. In this work, we continue to seek for another new but relatively simple
form for the neutral rotating squashed black hole solution by using a different metric ansatz of
time-like dimensional reduction. We then find its relation to our previous solution and investigate
its thermodynamics by means of the counterterm method. Compared with the previous results given by
the other author, both of our new metric forms and their associated thermodynamic expressions are
very concise and elegant. Both of two new forms for the neutral rotating squashed black hole solution
presented in this paper can be used as the seed to generate its charged generalization in $D=5$ minimal
supergravity.



\end{abstract}

\section{Introduction}\label{intr}

Ten years ago, Ishihara and Matsuno \cite{IM} obtained a new static charged solution of the squashed
black hole in the Einstein-Maxwell theory by applying the so-called squashing transformation to the
five-dimensional Reissner-Nordstr\"{o}m black hole solution. The method used by them is, in fact, a very
simple solution-generating technique with which some metric components of the known solution are multiplied
by different orders of a squashing function in order to get a new solution in the same theory. The solution
so obtained has a horizon topology of the squashed $S^3$ sphere while its asymptotical structure at spatial
infinity has the same asymptotic topology as that of the five-dimensional Kaluza-Klein (KK) magnetic monopole,
namely a twisted $U(1)$ fiber bundle over the four-dimensional Minkowski spacetime. That solution-generating
technique is also dubbed the name of squashing transformation method, and the obtained solution is usually
called as the KK squashed black hole because it becomes an exact solution in the four-dimensional KK theory,
after performing a dimensional reduction down to four dimensions. It belongs to a class of cohomogeneity-one
Kaluza-Klein black hole solutions \cite{TIr}.

Because of its simplicity and easy maneuverability, the squashing transformation method soon attracted a
great deal of attention and was then successfully applied to generate a lot of new squashed KK black hole
solutions \cite{WT,SSY,NY12,NY13,KN,NIMT,ST,SSW,MINT,TI,TIMN} from some known five-dimensional solutions.
In Ref. \cite{WT}, the squashing transformation was first applied to obtain a rotating squashed black
hole solution in vacuum gravity theory from the five-dimensional Myers-Perry black hole with two equal
rotation parameters. Subsequently, various squashed black hole solutions \cite{SSY,NY12,NY13,KN,NIMT,ST}
and those in the background of the G\"{o}del universe \cite{SSW,MINT,TI,TIMN} were constructed in the
Einstein-Maxwell-dilaton (EMd) theory \cite{SSY,NY12,NY13,KN}, the $D=5$ minimal supergravity theory
\cite{NIMT} and U(1)$^3$ supergravity theory \cite{ST}. On the other hand, in recent years there are also
many researches on various different aspects about squashed black hole solutions, such as thermodynamic
properties \cite{SSY,NY12,NY13,SSW,YN,NY11,CCO,KI07,KI08,SNT,PW}, Kerr/CFT correspondence \cite{PW},
geodetic motion \cite{MI}, Hawking radiation \cite{NY13,IS,CWS,WLLR,MU,LHL}, perturbation stability
\cite{KMIS,IKKMSZ,NK}, quasi-normal modes \cite{HWCCL,HWC}, and strong gravitational lens
\cite{LCJ,JCJ,SBV,CLJ,SNV}, etc.

However, the simplicity of applying the squashing procedure to get new exact black hole solutions will in
general give rise to much complexity in the analysis of their thermodynamic properties, since one has to
perform some further coordinate transformations to arrive at the proper asymptotical structure for the
metrics. Indeed, this is a very unpleasant matter for the charged rotating, squashed black hole solutions,
especially for those squashed counterparts of supergravity black holes with multiple electric charges.
In these cases, one can easily obtain the corresponding squashed black hole solutions through applying
a simple squashing transformation in the first step. But the resulting expressions of the solutions will
in general become very complicated when a further coordinate transformation is completed, and the expressions
calculated for the thermodynamic quantities are also very intricate. As such, thermodynamic properties
of the rotating charged, squashed black holes in the five-dimensional supergravity theory have not been
completely revealed so far in detail. What is more, for the rotating charged, squashed black holes, except
that the usual thermodynamic quantities (namely, the mass, the angular momentum, the electric charge, the
electrostatic potential, the Hawking temperature, the horizon entropy, the radii of the extra dimension
and the gravitational tension) have been taken into account, the first law of thermodynamics and the Smarr
mass formula still have to be modified \cite{SSW,PW} additionally via introducing the dipole potential and
the local dipole charge. This raises new issues on exploring thermodynamic properties of the rotating charged,
squashed black holes.

It is possible, from the very beginning, to obtain a relatively simple expression for the metric of the
rotating (charged) squashed black hole so that the subsequent analysis for its thermodynamic property is
fairly easy and the computed expressions for the thermodynamic quantities are also much more concise. This
is especially important for the case of a squashed black hole with multiple electric charges in supergravity
theory, since otherwise the asymptotic moduli of the dilaton scalar fields are nonzero at infinity and they
will further modify the first law of thermodynamics and the Smarr mass formula. If one can find a solution
of the squashed black hole with which the dilaton scalar fields vanish at infinity, this will greatly simplify
the analysis of black hole thermodynamics so that not only can the expressions calculated for the thermodynamic
quantities be very simple, but also one would not worry about any modification of first laws of thermodynamics
and the Smarr mass formula coming from the contribution of the asymptotic values of the dilaton scalar fields
at infinity. In fact, Yazadjiev \cite{SSY} had already considered this idea to obtain the static squashed KK
black hole solution in the EMd theory. Recently, this strategy was used in Ref. \cite{WWJY} to obtain a new
form for the five-dimensional static squashed black hole solution with three independent electric charges
that facilitates the analysis of its thermodynamic property, and the results indeed justify the validity
of this measure as expected.

However, the above-mentioned work \cite{SSY,WWJY} had only dealt with the static squashed black holes. Very
recently, as a warmup excise, we \cite{ZWWY} initiated the program to revisit the solution of a rotating
uncharged squashed black hole. By using the metric ansatz of space-like dimensional reduction and solving
directly the field equations of the vacuum Einstein gravity theory, we had gotten a new, simple form for
the five-dimensional neutral rotating squashed black hole solution and found that the expressions computed
for the thermodynamic quantities are also fairly concise. The neural rotating squashed solution found in
Ref. \cite{ZWWY} has two obvious advantages: (i) It is non-rotating at infinity; (ii) The Killing vector
along the time direction is properly normalized at infinity. Nevertheless, the big disappointing point is
that its $\tau\tau$-component (see Eq. (\ref{gtt}) below) of the metric is still very complicated, and our
motivation of this work is to overcome this shortcoming.

In this paper, we continue to look for another new but relatively simple form for the rotating uncharged
black hole with squashed horizons by using a different metric ansatz of time-like dimensional reduction.
Compared with our previous solution \cite{ZWWY}, the $tt$-component of our metric obtained in this time
has a particularly simple form although its time-like Killing vector is not properly normalized. We then
find its relation to the solution given in Ref. \cite{ZWWY} and investigate its thermodynamics by utilizing
the counterterm method \cite{MS}. Compared with the previous results given in Ref. \cite{WT}, both of our
new metric forms and their associated thermodynamic expressions of the neutral rotating squashed black
hole solution are very concise and elegant. We hope to further extend our experience to the case of the
charged generalization in the future work where two different forms of the solutions can be used as the
seed to generate the expected rotating charged squashed black hole.

The remaining parts of this article are organized as follows: Section \ref{s2} consists of the main body of
this work. To begin with, we first summarize briefly the main result in Ref. \cite{ZWWY} where a new form of
the five-dimensional neutral rotating squashed black hole solution was found by solving the vacuum Einstein
field equations with the assumption that the metric ansatz has the form of dimensional reduction along the
fifth spatial dimension. Then we will adopt a different metric ansatz of time-like dimensional reduction
to solve straightforwardly the vacuum field equation and find another new but relatively simple form for
the neutral rotating squashed black hole solution. Subsequently, we will establish its relation to the
previous one given in Ref. \cite{ZWWY}. In Section \ref{s3}, the counterterm method \cite{MS} is applied
to reveal its thermodynamic property. A brief summary and our future plan are given in the last section.
Appendix A includes the coordinate transformations and the necessary parameter identifications to relate
our solution presented in the context to that given in Ref. \cite{WT}. Appendix B establishes the relation
of our new form given in \cite{ZWWY} to the Dobiasch-Maison's solution \cite{DMP,GW}.

\section{New forms of the five-dimensional rotating squashed black hole}\label{s2}

As mentioned before, the rotating uncharged squashed black hole solution was first obtained in Ref. \cite{WT}
via applying the squashing transformation to the five-dimensional Myers-Perry black hole solution with two
equal rotation parameters. After further performing the suitable coordinate transformations, the expression
for the line element then becomes very involved, however, its asymptotic structure at spatial infinity, which
is the common behavior shared by all five-dimensional squashed black holes, is rather simple and is given
below in a frame non-rotating at infinity,
\be
ds^2 \simeq -d\tau^2 +d\rho^2 +\rho^2\big(d\theta^2 +\sin^2\theta\, d\phi^2\big) +\tL_\infty^2\sigma_3^2 \, ,
\ee
where $\sigma_3 = d\psi +\cos\theta\, d\phi$, and $\tL_\infty$ is the radii of the compact fifth dimension.

According to the above asymptotic behavior of the general five-dimensional squashed KK black hole, the action
reads
\be
I = \frac{1}{16\pi}\int d^5x \sqrt{-g}R +\frac{1}{8\pi}\int d^4x \sqrt{-h}K
 +\frac{1}{8\pi}\int d^4x\sqrt{-h}\sqrt{2\mathcal{R}}\, , \label{action}
\ee
in which the first term is the Einstein-Hilbert-Palatini action in five dimensions, $R$ is the Ricci scalar
corresponding to the five-dimensional metric $g_{\mu\nu}$. The second and third term are, respectively, the
Gibbons-Hawking boundary term and the counterterm proposed in Ref. \cite{MS}, $K$ is the trace of the extrinsic
curvature $K_{ij} = (n_{i;j} +n_{j;i})/2$ for the boundary with the induced metric $h_{ij}$, and $\mathcal{R}$
is the scalar curvature associated with the boundary metric $h_{ij}$.

\subsection{A new form for the rotating squashed black hole solution}\label{ss21}

In our recent work \cite{ZWWY}, a new, very simple form of the five-dimensional rotating uncharged squashed
black hole solution was obtained by solving the vacuum Einstein field equations. With the assumption that the
metric ansatz has the form of space-like dimensional reduction, we had solved directly the field equations
of the five-dimensional vacuum gravity theory to find the solution under the requirements that the solution
should satisfy two primary conditions: (i) It is non-rotating at spatial infinity; (ii) The time-like Killing
vector $\p_{\tau}$ is properly normalized at infinity.

For our purpose, the new form \cite{ZWWY} for the neutral rotating squashed black hole solution which is
non-rotating at infinity is given below (after a slight change of notations for no confusion in this paper)
\bea
&& d\tilde{s}^2 = -\,\frac{V(\rho)}{\widetilde{k}(\rho)}\, d\tau^2
 +\frac{\rho(\rho+\rho_0)}{V(\rho)}\, d\rho^2
 +\rho(\rho +\rho_0)\big(d\theta^2 +\sin^2\theta\, d\phi^2\big) \nn \\
&&\qquad\quad +\frac{\widetilde{k}(\rho)}{\rho(\rho +\rho_0)}\Big[\tL_\infty\sigma_3
 -\frac{\widetilde{h}(\rho)}{\widetilde{k}(\rho)}\, d\tau\Big]^2 \, , \label{os1}
\eea
where $\tL_\infty^2 = \rho_0^2 +\rho_0(\rho_+ +\rho_-) +2\rho_+\rho_-$, and
\bea
&& V(\rho) = (\rho -\rho_+)(\rho -\rho_-) \, , \qquad
 \widetilde{k}(\rho) = \rho^2 +\frac{\rho_0\rho_+\rho_-}{\tL_\infty^2}(2\rho +\rho_0) \, , \nn \\
&& \widetilde{h}(\rho) = \frac{\sqrt{\rho_+\rho_-(\rho_0 +\rho_+)(\rho_0 +\rho_-)(\tL_\infty^2
 -\rho_0^2)}}{\tL_\infty^2}(2\rho +\rho_0) \, . \nn
\eea
The above solution can also be written in the following form of time-like dimensional reduction
\bea
&& d\tilde{s}^2 = -\widetilde{f}(\rho)\bigg[d\tau
 +\frac{\widetilde{h}(\rho)\tL_\infty}{\rho(\rho +\rho_0)\widetilde{f}(\rho)}\,\sigma_3\bigg]^2
 +\frac{\rho(\rho +\rho_0)}{V(\rho)}\, d\rho^2 \nn \\
&&\qquad\quad +\rho(\rho +\rho_0)(d\theta^2 +\sin^2\theta\, d\phi^2)
 +\frac{V(\rho)\tL_\infty^2}{\rho(\rho +\rho_0)\widetilde{f}(\rho)}\,\sigma_3^2 \, , \label{os2}
\eea
in which the $\tau\tau$-component of the metric
\be\label{gtt}
\widetilde{f}(\rho) = \frac{\tL_\infty^2\rho^2 +\rho_0^3\rho -[(\rho_+ +\rho_-)\rho
 +(\rho_0 +\rho_+)(\rho_0 +\rho_-)](\tL_\infty^2 -\rho_0^2)}{\tL_\infty^2\rho(\rho +\rho_0)} \, .
\ee
is still very complicated although it approaches unity when $\rho\to \infty$, showing that the time-like
Killing vector $\p_{\tau}$ is properly normalized.

Due to the complicated form of the function $\widetilde{f}(\rho)$, once the above solution is used as the
seed to generate the charged version, the generated solution then will be very involved. This is the main
disappointing point of the above metric. It is this shortcoming that motivates us again to manage to look
for another new but relatively simple form for the rotating uncharged black hole with squashed horizons by
solving straightforwardly the vacuum Einstein field equations.

\subsection{Another new form for the rotating squashed black hole solution}\label{ss22}

In this subsection, we will adopt a different metric ansatz from that previously used in Ref. \cite{ZWWY} to
obtain another new but still relatively simple form for the neutral rotating black hole solution by solving
directly the vacuum Einstein field equations. The metric is assumed to take the form of time-like dimensional
reduction, which almost resembles the one suggested in Ref. \cite{Wu},
\be\begin{split}
ds^2 &= -f(\rho)\Big[dt +\frac{h(\rho)}{f(\rho)}\,\widetilde{\sigma}_3\Big]^2
 +\frac{L_\infty^2 V(\rho)}{\rho(\rho +\rho_0)f(\rho)}\,\widetilde{\sigma}_3^2 \\
&\quad +\frac{\rho(\rho +\rho_0)}{V(\rho)}\, d\rho^2
 +\rho(\rho +\rho_0)(d\theta^2 +\sin^2\theta\, d\phi^2) \\
& = \eta_{ab}e^a\otimes e^b \, ,
\end{split}\label{ms1}\ee
where $\widetilde{\sigma}_3 = d\widetilde{\psi} +\cos\theta d\phi$, and $\eta_{ab} = {\rm diag}(-1, 1, 1, 1, 1)$
is the Lorentzian metric in the orthogonal pentad frames.

Below we will solve the vacuum Einstein field equations within the f\"{u}nbein formalism. In order to simplify
the solving procedure as possible as we could, we choose the following pentad one-forms
\bea
&& e^1 = \sqrt{f(\rho)}\Big[dt +\frac{h(\rho)}{f(\rho)}\,\widetilde{\sigma}_3\Big]\, , \qquad
 e^2 = \frac{\sqrt{\rho(\rho +\rho_0)}}{\sqrt{V(\rho)}}\, d\rho \, , \nn \\
&& e^3 = \sqrt{\rho(\rho +\rho_0)}\, d\theta \, , \qquad\qquad\quad
 e^4 = \sqrt{\rho(\rho +\rho_0)}\sin\theta\, d\phi \, , \nn \\
&& e^5 = \frac{L_\infty\sqrt{V(\rho)}}{\sqrt{\rho(\rho
 +\rho_0)}\sqrt{f(\rho)}}\,\widetilde{\sigma}_3 \, , \nn
\eea
and have obviously $G_{(3)(3)} = G_{(4)(4)}$ for the pentad components of the vacuum Einstein field equations.
Our task below is to find the concrete expressions of three unknown functions $V(\rho)$, $f(\rho)$, and $h(\rho)$
via solving the equations of motion.

Firstly, from the composition: $G_{(2)(2)} +G_{(3)(3)} = 0$, one obtains the following equation
\be
\frac{\p^2}{\p\rho^2}V(\rho) = 2 \, , \label{G2p3}
\ee
which can be solved easily as
\be
V(\rho) = \rho^2 -v_1\rho +v_2 \,  , \label{Vs}
\ee
where $v_1$ and $v_2$ are two integration constants to be determined.

Secondly, the pentad component $G_{(1)(5)} = 0$ leads to the following relation
\be
\frac{\p}{\p\rho}\Big[\rho(\rho +\rho_0)\frac{\p f(\rho)}{\p\rho}\Big]
 +\frac{L_\infty^2}{\rho(\rho +\rho_0)}
 =  \frac{f(\rho)}{h(\rho)}\frac{\p}{\p\rho}\Big[\rho(\rho
 +\rho_0)\frac{\p h(\rho)}{\p\rho}\Big] \, , \label{G15}
\ee
while the composition: $G_{(2)(2)} -G_{(5)(5)} = 0$ results in a single differential equation that determines
the function $f(\rho)$ only
\be
\frac{\p}{\p\rho}\Big[\rho(\rho +\rho_0)\frac{\p f(\rho)}{\p\rho}\Big]
 +\frac{L_\infty^2 -\rho_0^2 f(\rho)}{\rho(\rho +\rho_0)} = 0 \, . \label{dfe}
\ee
Substituting Eq. (\ref{dfe}) into Eq. (\ref{G15}), one then gets the differential equation associated with the
function $h(\rho)$ only
\be
\frac{\p}{\p\rho}\Big[\rho(\rho +\rho_0)\frac{\p h(\rho)}{\p\rho}\Big]
= \frac{\rho_0^2h(\rho)}{\rho(\rho +\rho_0)} \, . \label{dhe}
\ee

The functions $f(\rho)$ and $h(\rho)$ can now be solved from Eqs. (\ref{dfe}) and (\ref{dhe}) by using the
Maple command `\emph{dsolve}' and their general expressions can be given as
\bes\bea
&& f(\rho) = \frac{L_\infty^2 -f_1\rho_0}{\rho_0^2} +\frac{f_2 -f_1}{\rho} +\frac{f_2}{\rho +\rho_0} \, , \\
&& h(\rho) = \frac{h_2\rho^2 +h_1(2\rho +\rho_0)\rho_0}{\rho(\rho +\rho_0)} \, , \label{hs}
\eea\ees
where ($f_1, f_2$) and ($h_1, h_2$) are four constants introduced in the process of integration.

From the consideration of the asymptotic property of the metric at infinity and the requirement of the
simplicity of the function $f(\rho)$ (otherwise, the final solution would be very complicated), we can
take it simply as
\be
f(\rho) = 1 -\frac{2m}{\rho} \, , \label{fs}
\ee
which gives $f_1 = 2m$, $f_2 = 0$, and
\be
L_\infty^2 = \rho_0^2 +2m\rho_0 \, .
\ee

By the self-consistence of all the field equations, we finally consider the composition of the pentad
components: $G_{(1)(1)} -G_{(2)(2)} +G_{(3)(3)} -G_{(5)(5)} = 0$, which gives the simplest relation
\bea
&& \Big[-(2\rho +\rho_0)\frac{\p V(\rho)}{\p\rho} +2V(\rho) +2\rho(\rho +\rho_0)\Big]f(\rho) \nn \\
&&\qquad\qquad\qquad +\frac{\rho_0^2 f(\rho) -L_\infty^2}{\rho(\rho +\rho_0)}V(\rho) +h(\rho)^2 = 0 \, .
\label{G1m2p3m5}
\eea
Substituting Eqs. (\ref{Vs}), (\ref{hs}) and (\ref{fs}) into Eq. (\ref{G1m2p3m5}), we can determine the
remaining four unknown integration constants as
\be
h_2 = h_1 \, , \quad v_1 = 2m -\frac{\rho_0 +m}{m\rho_0}h_1^2 \, ,
\quad v_2 = \frac{\rho_0}{2m}h_1^2 \, .
\ee
At this step, it is easy to check that the above solution can ensure that all the components of the vacuum
field equations are indeed completely satisfied.

Finally, we might set $h_1 = 2ma$ as well so that we can further obtain the following simplified expressions
for the solution
\be\begin{split}
& h(\rho) = 2ma\,\frac{\rho +\rho_0}{\rho} \, , \qquad f(\rho) = 1 -\frac{2m}{\rho} \, , \\
& V(\rho) = \rho^2 -2m\rho +2ma^2\,\frac{2(\rho_0 +m)\rho +\rho_0^2}{\rho_0} \, .
\end{split}\label{hfV}\ee
Thus, we have finished the solving process and obtained another new simple form for the rotating
neutral black hole with squashed horizons. The solution is obviously given by the line element (\ref{ms1})
with the structure functions (\ref{hfV}). Compared with the previous one given in Ref. \cite{WT}, the
solution presented here is much more simple.

By the way, our solution can also be put into the form of space-like dimensional reduction
\bea
&& ds^2 = -\,\frac{V(\rho)L_\infty^2}{k(\rho)}\, dt^2 +\frac{\rho(\rho+\rho_0)}{V(\rho)}\, d\rho^2
 +\rho(\rho +\rho_0)\big(d\theta^2 +\sin^2\theta\, d\phi^2\big) \nn \\
&&\qquad +\frac{k(\rho)}{\rho(\rho +\rho_0)}\Big[\widetilde{\sigma}_3
 -\frac{2ma(\rho+\rho_0)^2}{k(\rho)}\, dt\Big]^2 \, , \label{ms2}
\eea
where
\be
k(\rho) = \big(L_\infty^2 -4m^2a^2\big)\rho^2 +2ma^2\rho_0^2(2\rho+\rho_0) \, .
\ee

\subsection{Relation to our previous solution \cite{ZWWY}}\label{ss23}

In order to make contact with our previous one presented in the section \ref{ss21}, it is easy to observe
that the solution given in the last subsection has the following asymptotic behavior at infinity
\be
ds^2 \simeq -(dt +2ma\,\widetilde{\sigma}_3)^2 +L_\infty^2\widetilde{\sigma}_3^2 +d\rho^2
 +\rho^2\big(d\theta^2 +\sin^2\theta\, d\phi^2\big) \, ,
\ee
and one can see that the black hole is rotating at spatial infinity and the radii of the extra fifth dimension,
$\tL_\infty$, is given by
\be
\tL_\infty = \sqrt{L_\infty^2 -4m^2a^2} = \sqrt{\rho_0^2 +2m\rho_0 -4m^2a^2} \, .
\ee
To make the black hole non-rotating at infinity, it needs to perform the following coordinate transformations
\be
t = \frac{\tL_\infty}{L_\infty}\tau \, , \quad \widetilde{\psi} = \psi +\frac{2ma}{L_\infty\tL_\infty}\tau \, ,
\label{cts}
\ee
and now the metric (\ref{ms1}) takes the following form
\bea
&& ds^2 = -f(\rho)\bigg[\frac{\tL_\infty}{L_\infty}d\tau
 +\frac{h(\rho)}{f(\rho)}\Big(\sigma_3 +\frac{2ma}{L_\infty\tL_\infty}d\tau \Big)\bigg]^2 \nn \\
&&\qquad\quad  +\frac{\rho(\rho +\rho_0)}{V(\rho)}\, d\rho^2
 +\rho(\rho +\rho_0)(d\theta^2 +\sin^2\theta\, d\phi^2) \nn \\
&&\qquad\quad +\frac{L_\infty^2V(\rho)}{\rho(\rho +\rho_0)f(\rho)}\Big(\sigma_3
 +\frac{2ma}{L_\infty\tL_\infty}d\tau \Big)^2 \, . \quad \label{ms3}
\eea

To equate the solution (\ref{ms3}) with (\ref{os1}) [or (\ref{ms2}) after using Eq. (\ref{cts})],
it remains only necessary to make the following parameter identifications:
\bes\bea
&& m = \rho_0\frac{\rho_0(\rho_+ +\rho_-) +2\rho_+\rho_-}{2(\rho_0^2 -\rho_+\rho_-)} \, , \label{relm} \\
&& a^2 = \frac{\rho_+\rho_-(1 -\rho_+\rho_-/\rho_0^2)}{\rho_0(\rho_+ +\rho_-) +2\rho_+\rho_-} \, . \label{rela}
\eea\ees

\section{Thermodynamics}\label{s3}

In this section, we will investigate thermodynamic property of the neutral rotating squashed black hole
based upon the solution (\ref{ms3}). The locations of the inner and outer horizons are determined by
$V(\rpm) = 0$, which can be explicitly expressed in terms of the parameters ($m, a, \rho_0$) but will be
omitted here. It is a standard excise to obtain the entropy $S_{\pm} = A_{\pm}/4$, Hawking temperature
$T_{\pm} = \kappa_{\pm}/(2\pi)$ and the angular velocity $\Omega_{\pm}$ on the horizons as follows:
\bes\bea
&& S_{\pm} = 4\pi^2\frac{\rpm(\rpm +\rho_0)h(\rpm)}{\sqrt{-f(\rpm)}} \, , \\
&& T_{\pm} = \frac{\tL_\infty\sqrt{-f(\rpm)}}{4\pi\rpm(\rpm +\rho_0)h(\rpm)}\cdot
 \frac{\p V(\rpm)}{\p\rpm} \, , \\
&& \Omega_{\pm} = -\, \frac{f(\rpm)\tL_\infty^2 +2ma h(\rpm)}{L_\infty\tL_\infty\, h(\rpm)} \, .
\eea\ees

Our next task is to calculate the conserved charges: the counterterm mass, the angular momentum and the
gravitational tension via the counterterm method \cite{MS}. Varying the action (\ref{action}) with the
induced metric $h_{ij}$ leads to the following boundary stress-energy tensor
\be
8\pi T_{ij} = K_{ij} -h_{ij}K -\Psi(\mathcal{R}_{ij} -h_{ij}\mathcal{R})
-h_{ij}h^{kl}\Psi_{;kl} +\Psi_{;ij} \, ,
\ee
where $\Psi = \sqrt{2/\mathcal{R}}$. After some tedious calculations, we obtain the following asymptotic
expansion of the coordinate components of the stress tensor
\bes\bea
&& T^\tau_{~~\tau} = -\, \frac{(\tL_\infty^2 -4ma^2\rho_0)(2\tL_\infty^2
 -\rho_0^2)}{16\pi\tL_\infty^2\rho_0\rho^2} +\mO(\rho^{-3}) \, ,\label{tt} \\
&& T^t_{~~\psi} = \frac{ma(\rho_0-2ma^2)L_\infty}{4\pi\tL_\infty\rho^2} +\mO(\rho^{-3}) \, , \label{tp} \\
&& T^{\psi}_{~~\psi} = -\, \frac{(\tL_\infty^2 -4ma^2\rho_0)(\tL_\infty^2
 -m\rho_0 +2m^2a^2)}{8\pi\tL_\infty^2\rho_0\rho^2} +\mO(\rho^{-3}), \label{pp} \\
&& T^\phi_{~~\phi} = \frac{4ma^2(\rho_0+m)^2 -\rho_0(2L_\infty^2 -m\rho_0+4m^2)}{16\pi\rho_0^2\rho^3}ma^2 \nn \\
&&\qquad\quad -\,\frac{L_\infty^2-4m^2}{64\pi\rho^3} +\mO(\rho^{-4}) \, , \\
&& T^{\psi}_{~~\tau} \approx -T^\tau_{~~\psi}/\tL_\infty^2 \, , \qquad
T^\phi_{~~\psi} = T^\phi_{~~\tau} = 0 \, , \\
&& T^\tau_{~~\phi} = T^\tau_{~~\psi}\cos\theta \, , \qquad
T^{\psi}_{~~\phi} \approx T^{\psi}_{~~\psi}\cos\theta \, .
\eea\ees

Now we can first calculate the counterterm mass and the angular momentum via the following formulae
\bes\bea
&&\hspace*{-0.3cm} M_{ct} = \frac{-1}{8\pi}\int_0^{2\pi}d\phi \int_0^{4\pi}d\psi \int_0^\pi d\theta\,
 \big(\!\sqrt{\Sigma}T^t_{~~t}\big)\big|_{\rho\to\infty} \, , \\
&& J = \frac{-1}{8\pi}\int_0^{2\pi}d\phi \int_0^{4\pi}d\psi \int_0^\pi d\theta\,
 \big(\!\sqrt{\Sigma}T^t_{~~\psi}\big)\big|_{\rho\to\infty} \, ,
\eea\ees
where
$$\sqrt{\Sigma} = \sqrt{\frac{L_\infty^2V(\rho) -\rho(\rho+\rho_0)h(\rho)^2}{f(\rho)}}
\sqrt{\rho(\rho+\rho_0)}\sin\theta \, .$$

After substituting the expressions (\ref{tt},\ref{tp}) into the above two formulae, the expressions
of the counterterm mass and the angular momentum can be computed as
\bes\be\begin{split}
M_{ct} &= \frac{\pi(2\tL_\infty^2 -\rho_0^2)(\tL_\infty^2-4ma^2\rho_0)}{\tL_\infty\rho_0} \\
&= \frac{\pi(\rho_0 +\rho_+ +\rho_-)(\rho_0 +2\rho_+)(\rho_0 +2\rho_-)}{\tL_\infty} \, ,
\end{split}\ee
\be\begin{split}
J &= 4\pi ma(\rho_0 -2ma^2)L_\infty \\
&= 2\pi\sqrt{\rho_+\rho_-(\rho_0 +\rho_+)(\rho_0 +\rho_-)(\tL_\infty^2 -\rho_0^2)} \, .
\end{split}\ee\ees
Note that the angular momentum is identical to the result obtained by using the Komar method, while the
counterterm mass is different from the Komar mass.

Then the non-zero gravitational tension can be computed by using the following formula
\be
\mathcal{T} = \frac{-1}{8\pi}\int_0^{2\pi}d\phi \int_0^\pi d\theta\,
 \big(\!\sqrt{\sigma}T^{\psi}_{~~\psi}\big)\big|_{\rho\to\infty} \, ,
\ee
where $\sqrt{\sigma} = \rho(\rho +\rho_0)\sin\theta$. Using Eq. (\ref{pp}), we can get the expression of
the gravitational tension as follows:
\be
\mathcal{T} = \frac{(\tL_\infty^2 -4ma^2\rho_0)(\tL_\infty^2 +\rho_0^2)}{4\tL_\infty^2\rho_0}
 = \frac{(\rho_0 +\rho_+ +\rho_-)(\tL_\infty^2 +\rho_0^2)}{4\tL_\infty^2} \, .
\ee

It can be checked that the above thermodynamic quantities are essentially identical to those previously
given in Ref. \cite{ZWWY} via the relations (\ref{relm}, \ref{rela}). Compared with the results in Ref.
\cite{WT}, our thermodynamical expressions are much more concise.

Finally, it is not difficult to verify that the above thermodynamical quantities completely fulfil both
of the differential and the integral first laws of black hole thermodynamics
\bes\bea
&&\hspace*{-0.2cm} dM_{ct} = T_{\pm}\, dS_{\pm}  +\Omega_{\pm}\, dJ  +4\pi\mathcal{T}\, d\tL_\infty \, , \\
&& M_{ct} = 3T_{\pm} S_{\pm}/2 +\Omega_{\pm} J +2\pi\mathcal{T} \tL_\infty \, .
\eea\ees
where $2\pi\tL_\infty$ is the length of the compact fifth dimension, and $\tL_\infty$ can be roughly
identified with twice of the NUT charge, which is viewed here as a thermodynamical variable for the
self-consistence of the Smarr mass formula.

\section{Conclusion}

In this paper, we have obtained another new and rather simple form for the five-dimensional neutral
rotating squashed KK black hole by adopting a different metric ansatz to solve directly the vacuum
Einstein field equations. The black hole is rotating at spatial infinity, but when transformed into
a frame non-rotating at infinity, it can be cast into the form previously found in Ref. \cite{ZWWY}
if the parameter identifications (\ref{relm}, \ref{rela}) are further made. Obviously, our new metric
expression for the rotating squashed black hole is much simpler than the previous one given in Ref.
\cite{WT} and is very convenient for us to investigate its thermodynamic property. Then the counterterm
method has been used to compute its conserved charges: the counterterm mass, the angular momentum and
the gravitational tension, and it has been verified that they completely satisfy the differential first
law and the Smarr mass formula.

It should be emphasized that compared with the previous research \cite{WT}, not only is our new metric
form present here or that in Ref. \cite{ZWWY} for the neutral rotating squashed black hole much more
simple, but also its associated thermodynamic expressions are very concise and elegant. We think that
the new form of the neutral rotating squashed black hole presented in this paper is the most perfect
seed to generate its charged generalization in the near future. By the way, we have also provided two
appendices to establish the relations of our new expressions to the previous one given in Ref. \cite{WT}
and the Dobiasch-Maison's solution \cite{DMP,GW}.

\section*{Acknowledgement}
This work is supported in part by the NSFC under Grant Nos. 10975058, 11275157, 11675130 and 11573022.

\appendix
\section{Relation to the previous solution \cite{WT}}
\label{appa}

\setcounter{equation}{0}
\renewcommand{\theequation}{A.\arabic{equation}}

In this appendix, we will show that how our solution (\ref{ms1}) with the structure functions (\ref{hfV})
can be obtained from that given in Ref. \cite{WT} via the coordinate transformations and the appropriate
parameter identifications. Making a coordinate shift $r^2 +\hat{a}^2\to r^2$ and simultaneously, $r_\infty^2
+\hat{a}^2 \to r_\infty^2$, the neutral rotating squashed KK black hole solution \cite{WT} can be written
into a form that resembles closely the one adopted previously in Ref. \cite{Wu}
\bea
&& d\hat{s}^2 = -\hat{f}(r)\Big[d\hat{t} +\frac{\hat{h}(r)}{\hat{f}(r)}\,\widetilde{\sigma}_3\Big]^2
 +\frac{r^2\hat{V}(r)}{4\hat{f}(r)}\,\widetilde{\sigma}_3^2 +\frac{k(r)^2}{\hat{V}(r)}\, dr^2 \nn \\
&&\qquad\quad +\frac{r^2k(r)}{4}(d\theta^2 +\sin^2\theta\, d\phi^2) \, ,
\eea
where $\widetilde{\sigma}_3 = d\widetilde{\psi} +\cos\theta\, d\phi$, and
\bea
&& \hat{f}(r) = 1 -\frac{2\hat{m}}{r^2} \, , \qquad
\hat{V}(r) = 1 -\frac{2\hat{m}}{r^2} +\frac{2\hat{m}\hat{a}^2}{r^4} \, , \nn \\
&& \hat{h}(r) = \frac{\hat{m}\hat{a}}{r^2} \, , \qquad\quad~~~
k(r) = \frac{\hat{V}(r_\infty)}{(1 -r^2/r_\infty^2)^2} \, . \nn
\eea

If we would like to transform it into the form (\ref{ms1}), namely,
\bea
&& ds^2 = -f(\rho)\Big[dt +\frac{h(\rho)}{f(\rho)}\,\widetilde{\sigma}_3\Big]^2
 +\frac{L_\infty^2\rho\bar{V}(\rho)}{(\rho +\rho_0)f(\rho)}\,\widetilde{\sigma}_3^2 \nn \\
&&\qquad\quad +\frac{\rho +\rho_0}{\rho\bar{V}(\rho)}\, d\rho^2
 +\rho(\rho +\rho_0)(d\theta^2 +\sin^2\theta\, d\phi^2) \, ,
\eea
we must first make the following coordinate transformations
\be
\frac{r^2}{r_\infty^2} = \frac{\rho}{\rho +\rho_0} \, , \qquad
t = \hat{t}\sqrt{\hat{f}(r_\infty)} \equiv \frac{\rho_0}{L_\infty}\, \hat{t} \, ,
\ee
and meanwhile the parameter identifications:
\be
\rho_0^2 = \frac{r_\infty^2\hat{V}(r_\infty)}{4} \, , \qquad
L_\infty^2 = \frac{r_\infty^2\hat{V}(r_\infty)}{4\hat{f}(r_\infty)}
 = \frac{\rho_0^2}{\hat{f}(r_\infty)} \, ,
\ee
in which
\bea
\hat{f}(r_\infty) = 1 -\frac{2\hat{m}}{r_\infty^2} \, , \qquad
\hat{V}(r_\infty) = 1 -\frac{2\hat{m}}{r_\infty^2} +\frac{2\hat{m}\hat{a}^2}{r_\infty^4} \, . \nn
\eea

Using the identity $\hat{f}(r_\infty) = \rho_0^2/L_\infty^2$, we then obtain
\be
f(\rho) = \frac{\hat{f}(r)}{\hat{f}(r_\infty)}
 = 1 -\frac{2\hat{m}\rho_0}{\hat{f}(r_\infty) r_\infty^2\rho}
 = 1 -\frac{2\hat{m}L_\infty^2}{r_\infty^2\rho_0\rho}
\equiv 1 -\frac{2m}{\rho} \, ,
\ee
where in the above we have set
\be
\frac{\hat{m}}{r_\infty^2} = \frac{m\rho_0}{L_\infty^2} \, ,
\ee
for the sake of the simplicity of the function $f(\rho)$. Then, in order to ensure that the following
identity
\be
\hat{f}(r_\infty) = 1 -\frac{2m\rho_0}{L_\infty^2} = \frac{\rho_0^2}{L_\infty^2} \, ,
\ee
is satisfied, one must let
\be
L_\infty^2 = \rho_0^2 +2m\rho_0 \, .
\ee

Next, one can show that
\be
h(\rho) = \frac{\hat{h}(r)}{\sqrt{\hat{f}(r_\infty)}}
 = \frac{\hat{m}\hat{a}L_\infty(\rho +\rho_0)}{r_\infty^2\rho_0\rho}
 = \frac{m\hat{a}(\rho +\rho_0)}{L_\infty\rho}
\equiv 2ma\frac{\rho +\rho_0}{\rho} \, ,
\ee
where to arrive at the last identity, one should further make another parameter identification
\be
\hat{a} = 2aL_\infty \, .
\ee

Finally, it is a little troublesome to get the expression of the function $\bar{V}(\rho)$. However,
using the above relations one can deduce that
\be
\hat{V}(r_\infty) = 1 -\frac{2m\rho_0}{L_\infty^2} +\frac{8ma^2\rho_0}{r_\infty^2}
= \frac{\rho_0^2}{L_\infty^2} +\frac{8ma^2\rho_0}{r_\infty^2}
\equiv \frac{4\rho_0^2}{r_\infty^2} \, . \nn
\ee
From this equation, one can get a useful identity:
\be
\frac{r_\infty^2}{4L_\infty^2} = 1 -\frac{2ma^2}{\rho_0} \, ,
\ee
and with which, one can finally show that
\bea
&&\bar{V}(\rho) = \frac{\hat{V}(r)}{\hat{V}(r_\infty)}
= \frac{r_\infty^2(\rho -2m)}{4L_\infty^2\rho} +2ma^2\frac{(\rho+\rho_0)^2}{\rho_0\rho^2} \nn \\
&&\quad~~~ = 1 -\frac{2m}{\rho} +2ma^2\frac{2(\rho_0+m)\rho +\rho_0^2}{\rho_0\rho^2} \, .
\eea

\section{Relation to the Dobiasch-Maison's solution \cite{DMP,GW}}
\label{appb}

\setcounter{equation}{0}
\renewcommand{\theequation}{B.\arabic{equation}}

In Ref. \cite{TIr}, there are two baffling statements that ``It was shown by Wang$^{51)}$ that the
five-dimensional Kaluza-Klein black hole of Dobiasch and Maison can be reproduced by squashing a
five-dimensional Myers-Perry black hole with two equal angular momenta." and ``(The Dobiasch-Maison
solution was re-derived by squashing from the five-dimensional Myers-Perry with equal angular momenta
$^{51)}$)". However, Wang neither showed this reproduction in his paper \cite{WT} nor referred to the
Dobiasch-Maison's work \cite{DMP}.

In the last appendix, we have demonstrated that our solution can be transformed from Wang's solution
\cite{WT}. It is therefore also meaningful to establish here its relation to the Dobiasch-Maison's
solution \cite{DMP,GW}.

\subsection*{B.1 The Dobiasch-Maison's solution \cite{DMP,GW}}

Dobiasch and Maison \cite{DMP} found a three-parameter solution in the five-dimensional vacuum gravity
theory. It was later studied \cite{GW} in detail and was shown to describe a static dyonic KK black hole
after performing a dimension reduction down to four-dimensions. Its original form \cite{DMP} can be written
as
\bea
&& ds^2 = -\, \frac{\Delta}{B}\, d\tau^2 +A\Big(\frac{dr^2}{\Delta} +d\theta^2 +\sin^2\theta\, d\phi^2\Big) \nn \\
&&\qquad\qquad +\frac{B}{A}\Big(dy +\frac{2\delta\,r}{B}\, d\tau +2\gamma\cos\theta\, d\phi\Big)^2 \, ,
\label{dmgw1}
\eea
where
\bea
&& A = r^2 -\frac{\alpha+\beta}{\alpha-\beta}\big(\alpha^2 -a^2\big) \, , \quad
B = (r +\alpha+\beta)^2 -\frac{\alpha+\beta}{\alpha-\beta}\big(a^2 -\beta^2\big) \, , \nn \\
&&\qquad \Delta = (r +\alpha)^2 -a^2 \, , \qquad
\gamma^2 = \frac{\alpha(\alpha^2 -a^2)}{\alpha-\beta} \, , \quad
\delta^2 = \frac{\beta(a^2 -\beta^2)}{\alpha-\beta} \, , \nn
\eea
in which three parameters ($a, \alpha, \beta$) are twice of those in Ref. \cite{DMP}.

By solving the equation of motion, we find that the above solution can also be given in terms of the
electric charge $\delta$, the magnetic charge $\gamma$, and the third parameter $a_0$ as follows:
\bea
&& A = r^2 -\gamma^2 -a_0 \, , \qquad
B = r^2 -2r\sqrt{\frac{(a_0+\gamma^2)(\delta^2+a_0)}{a_0}} +\gamma^2 +a_0 \, , \nn \\
&&\hspace{4.1cm} \Delta = r^2 -2r\gamma^2\sqrt{\frac{\delta^2+a_0}{a_0(a_0+\gamma^2)}} +\gamma^2 -a_0 \, . \nn
\eea

On the other hand, the above static dyonic KK black hole solution was given \cite{GW} in a form completely
expressed in terms of the physical parameters (the mass $M$, the dilaton scalar charge $\Sigma$, the electric
charge $Q$, and the magnetic charge $P$). Its rotating extension is the Rasheed-Larsen rotating dyonic black
hole \cite{RL}, which belongs to the cohomogeneity-two class of Kaluza-Klein black hole solutions \cite{TIr}.
After setting the rotation to zero, re-scaling $\tilde{\Sigma}\to \sqrt{3}\Sigma$, and then shifting $\tilde{r}
\to r +\Sigma$, the static dyonic KK black hole solution \cite{GW} (see also Ref. \cite{CGS}) is given as
\bea
&& ds^2 = -\, \frac{\Delta}{B}\, d\tau^2 +A\Big(\frac{dr^2}{\Delta} +d\theta^2 +\sin^2\theta\, d\phi^2\Big) \nn \\
&&\qquad\qquad +\frac{B}{A}\Big(dy +\frac{2Qr}{B}\, d\tau +2P\cos\theta\, d\phi\Big)^2 \, ,
\label{dmgw2}
\eea
where
\bea
&& A = r^2 -\frac{2P^2\Sigma}{\Sigma-M} \, , \qquad
 B = (r +2\Sigma)^2 -\frac{2Q^2\Sigma}{\Sigma+M} \, , \nn \\
&& \Delta = (r -M +\Sigma)^2 +P^2 +Q^2 -M^2 -3\Sigma^2 \, . \nn
\eea

The solution (\ref{dmgw1}) and (\ref{dmgw2}) are related to each other via the following relations
\bea
M = \frac{\beta -\alpha}{2} \, , \quad \Sigma = \frac{\alpha +\beta}{2}\, , \quad
P^2 = \frac{\alpha(\alpha^2 -a^2)}{\alpha-\beta} \, , \quad
Q^2 = \frac{\beta(a^2 -\beta^2)}{\alpha-\beta} \, . \quad
\eea
The mass $M$, the dilaton scalar charge $\Sigma$, the electric charge $Q = \delta$, and the magnetic charge
$P = \gamma$ are not all independent, rather they must satisfy a constraint condition
$$\frac{P^2}{\Sigma-M} +\frac{Q^2}{\Sigma+M} = 2\Sigma \, .$$

\subsection*{B2. Relation to the Dobiasch-Maison's solution \cite{DMP,GW}}

It is much more easy to establish the relation of our previous solution (\ref{os1}) presented in Ref.
\cite{ZWWY} to the Dobiasch-Maison's solution \cite{DMP,GW} given in the above expression (\ref{dmgw2}).
The coordinate transformations followed by the parameter identifications given below can do the job:
\bea
&& r = \rho +\rho_0/2 \, , \qquad y = \tL_\infty\psi \, , \qquad\qquad
\rho_0^2 = \frac{8P^2\Sigma}{\Sigma-M}\, , \nn \\
&& P = \tL_\infty/2 \, , \quad
Q = -\,\frac{\sqrt{\rho_+\rho_-(\rho_0+\rho_+)(\rho_0+\rho_-)(\tL_\infty^2 -\rho_0^2)}}{\tL_\infty^2} \, , \nn \\
&& M = \frac{\rho_0}{4} +\frac{\rho_+ +\rho_-}{2} +\frac{\rho_0\rho_+\rho_-}{2\tL_\infty^2} \, , \qquad
\Sigma = \frac{\rho_0}{4}\Big(\frac{2\rho_+\rho_-}{\tL_\infty^2} -1\Big) \, , \nn
\eea
where $\tL_\infty^2 = \rho_0^2 +\rho_0(\rho_+ +\rho_-) +2\rho_+\rho_-$.

\end{document}